\documentclass[prl,twocolumn,showpacs,floatfix,superscriptaddress]{revtex4}
\usepackage{graphicx,amsfonts,amssymb,amsmath, hyperref}

\newif\ifhyper
\hypertrue
\ifhyper
\hypersetup{
  citecolor = {green},
  colorlinks = {true}, % false
  urlcolor = {blue} % magenta
}
\fi

\newlength{\ldag}
\begin{document}

\title{Crumpling transition and flat phase of  polymerized  phantom membranes}

\author{J.-P.  Kownacki}
\affiliation{LPTM, CNRS UMR 8089, Universit\'e de Cergy-Pontoise,   2 Avenue Adolphe Chauvin, 95302 Cergy-Pontoise  Cedex, France}
\email{kownacki@u-cergy.fr}

\author{D. Mouhanna} 
\affiliation{LPTMC,
CNRS UMR 7600, UPMC, 4 Place Jussieu, 75252 Paris Cedex 05, France}
\email{mouhanna@lptmc.jussieu.fr}

%------------------------------------------------------------------------------

\begin{abstract}
Polymerized phantom membranes are  revisited using a nonperturbative renormalization group approach.  This allows one  to investigate  both  the crumpling transition  and   the low-temperature, flat,  phase   in any  internal  dimension $D$ and embedding dimension $d$, and   to determine  the lower critical dimension.    The crumpling phase transition for physical membranes is found to be  of second order  within our approximation. A    weak first-order behavior, as observed in recent Monte Carlo simulations, is however not excluded.

\end{abstract}


\maketitle

Membranes  form   a particularly  rich  and  exciting  domain   of statistical physics  in which  the  interplay between  two-dimensional geometry and thermal fluctuations  has led to  a lot of  unexpected  behaviors  going from flat to tubular and  glassy  phases (see \cite{proceedings89,radzihovsky04,bowick01,nelson04} for reviews).  Roughly speaking, membranes fall into two groups \cite{nelson04}: fluid membranes,   in which  the building monomers  are free to diffuse.   The  connectivity  is thus not fixed and the membrane  displays   a vanishing shear  modulus.  In  contrast,  in polymerized   membranes  the  monomers are tied together through a potential  which  leads to a fixed connectivity  and to elastic forces.  While  fluid membranes are always crumpled,  polymerized membranes, due to their nontrivial elastic properties,  exhibit a  phase transition between a crumpled phase at high temperature and a   flat phase at low temperature  with  orientational order between the normals  of the  membrane \cite{nelson87,david88,paczuski88,nelson04}.  Amazingly, due to  the existence of long-range forces mediated by phonons,  the correlation functions in the flat phase display a nontrivial  infrared scaling behavior \cite{aronovitz88,guitter89,aronovitz89}. Accordingly,    the lower critical dimension above which  an order can develop  appears    to be smaller than 2 \cite{aronovitz89},  in apparent violation of the Mermin-Wagner theorem.

Let us  consider   the general case of $D$-dimensional non self-avoiding (phantom) membranes embedded in a $d$-dimensional space.  Early   $\epsilon$-expansion \cite{paczuski88}  performed at one-loop order  on the  Landau-Ginzburg-Wilson-type model relevant to study the crumpling transition of  polymerized  membranes has led to   predict  that  just below  the upper critical dimension  $D=4$,  the crumpling  transition is of second order for  $d>d_{cr}= 219$     while   it is of  first order  for  $d<d_{cr}$. This  leaves however   open the question of the nature of  the transition  in the physical ($D=2, d=3$)  situation,  the  case  $\epsilon=2$   being clearly out of reach of  such a one-loop order computation.   On the  numerical  side  former Monte Carlo (MC)  studies    (see \cite{gompper04,kantor04} for reviews)    predict a second-order  behavior while more recent   simulations   \cite{kownacki02,koibuchi04}  rather  favor    first-order behaviors.   There is  however no definite conclusion  and  no  explanation for  these versatile results.

 In  parallel to  the investigation  of  the crumpling transition,  an effective  elastic field  theory   has been   used  to probe   the  flat, low-temperature,   phase of membranes \cite{nelson87,aronovitz88,aronovitz89,nelson04}.   An   $\epsilon$-expansion has  been performed \cite{aronovitz88},   also at one-loop order,  below the upper critical dimension  $D=4$ showing  that this  flat phase is controlled  by a  nontrivial fixed point  (FP).  However,  again, this low order computation performed in the  vicinity of  $D=4$  has been of no  use  to accurately  determine  the properties of  genuine  $2D$ membranes   such as   the  critical exponents  and  the lower critical dimension  $D_{lc}(d)$ above  which the  flat phase can  exist.  

Significant  progress  has   been realized with the use of  large-$d$ expansion \cite{paczuski89,aronovitz89},  and  variant of,  such as self-consistent screening approximations (SCSA)  \cite{ledoussal92}  that   have allowed  to evaluate    the exponent $\eta$  both at the crumpling transition and  in  the flat phase as well as   the dimension $D_{lc}(d)$. However, the very nature  of the approach, requiring  large values of  $d$,   makes  doubtful  the quantitative  predictions  extrapolated at small $d$  and even  impossible  the determination of     the   line  $d_{cr}(D)$,   separating  the first-order from the second-order regions.

A  flaw of the previous approaches  to   polymerized  membranes   is that,  due to  their  perturbative character,   they are     unable   to treat  all     aspects of the physics of membranes   including  crumpling transition,  flat phase,  and lower critical dimension and  thus  to get  a global  picture of the  renormalization  group  (RG)  phase diagram.  In this article,  we propose an   approach of polymerized membranes  based on  a nonperturbative RG  method \cite{wetterich93c} which has been applied  successfully in  both particle and condensed matter physics (see \cite{bagnuls01, berges02, delamotte03} for reviews).  With  this method, that we adapt  to  the treatment of extended  objects,   we  are  able to describe  within  the {\sl  same} formalism,  {\it i.e.}, using a {\sl   unique} effective action and a  {\sl  unique}  set of RG equations,  both the crumpling transition and the  flat phase of membranes.   Concerning   the crumpling transition,  we   reproduce  the results obtained  within the  $\epsilon$-expansion approach and at leading order within the large-$d$ approaches.    Moreover,  we  determine the line $d_{cr}(D)$  everywhere between $D=4$ and  $D=2$.  Our estimates   of  $d_{cr}(D=2)\simeq  2$ lead  to   predict   a  second-order phase transition  in the physical case but  do not completely exclude a   weak first-order behavior.    Our investigation of the flat phase  also  allows    to recover all previous perturbative results including $\epsilon$ and $1/d$-expansions.  Moreover we get, {\sl for all values of $d$},  a determination  of the lower critical dimension $D_{lc}(d)$   above which the crumpling transition 
and flat  phase fixed points  are shown to coexist.

Our approach is based  on the  concept   of  effective average action  \cite{wetterich93c} (see \cite{bagnuls01,berges02, delamotte03} for reviews), $\Gamma_k[\bf r]$,  where   $\bf r=r({\bf x})$  is    a $d$-dimensional   {\sl  external}  vector  that  describes the membrane  in the embedding space  while    ${\bf x}$   is   a  set of   {\sl  internal}  $D$-dimensional  coordinates  which  labels a point   within the membrane. The  quantity  $\Gamma_k[{\bf r}]$, $k$ being a running scale going from a lattice scale $k=\Lambda$ to the infrared scale  $k=0$,  has the physical meaning of  a coarse grained free energy where only fluctuations with momenta $q\ge k$  have been integrated out.   Thus, at  the lattice scale  $\Lambda$, $\Gamma_{k=\Lambda}$  identifies with the   continuum limit of  some  lattice Hamiltonian while   at long distance, { \it i.e.},  at $k=0$, it identifies with the  standard free energy $\Gamma$. The $k$-dependence, RG flow,  of $\Gamma_k$ is  provided  by an exact evolution equation \cite{wetterich93c},
\begin{equation}
{\partial \Gamma_k\over \partial t}={1\over 2} \hbox{Tr} \left\{(\Gamma_k^{(2)}+R_k)^{-1}
 {\partial R_k\over \partial t}\right\}
\label{renorm}
\end{equation}
where $t=\ln \displaystyle {k / \Lambda}$. The trace has to be understood as a $D$-dimensional momentum
 integral as well as a summation over internal indices. In Eq.(\ref{renorm}), $R_k(q)$ is an
 effective infrared cut-off function which suppresses  the propagation of modes with momenta $q<k$ and makes that $\Gamma_k$ encodes only modes with momenta $q\ge k$. A convenient cut-off is provided by $R_k(q)= Z  (k^4-q^4)\theta(k^2-q^2)$, where $Z$ is a field renormalization -- see below.  It generalizes to  translational-invariant  action density  a   cut-off  \cite{litim02} which  has been largely  used since it leads to compact expressions. Finally,  and of  utmost importance,  note that the  term  $\Gamma_k^{(2)}$  in Eq.(\ref{renorm}) is, in principle,   the exact,  {\it  i.e.}, {\sl full field-dependent},  inverse propagator, the second derivative of $\Gamma_k$ with respect to the field $\bf r$, taken in a {\sl generic}, nonvanishing field configuration. This is  the fact   at  the very   origin of the nonperturbative character  of the method.

Let us now make precise the form of    $\Gamma_k$.   It  must be invariant under  the group  of Euclidean displacements  which includes translations and rotations. This imposes to $\Gamma_k$ to be a functional of  $\partial_{\alpha}{\bf r}\equiv\partial{\bf r}/ \partial{x_\alpha}$, $\alpha=1 \dots D$,  the order parameter,  and of scalars  in both  the embedding  and     membrane spaces.   An exact treatment  of  Eq.(\ref{renorm})   would imply $\Gamma_k$  to  enclose    all   powers and derivatives of   these Euclidean invariants. This  goal is however   unrealistic  and  one  has  to truncate $\Gamma_k$. We choose here an ansatz    that   allows both  to make easily contact with  previous -- perturbative -- approaches   and  to   realize our program. It is given by:
\begin{eqnarray}
\displaystyle \Gamma_k \left(\bf{r}\right)=\int d^ D x\  &\displaystyle{Z\over 2}& \left(\partial_{\alpha}\partial_{\alpha}\bf{r}\right)^2  
+ u\left(\partial_{\alpha}{\bf r}. \partial_{\beta}{\bf r}-\zeta^2\delta_{\alpha\beta}\right)^2\nonumber\\
\label{action}
&+&\ v\left(\partial_{\alpha}{\bf r}. \partial_{\alpha}{\bf r}-D \zeta^2 \right)^2
\end{eqnarray} 
where ${Z, u,v}$  and $\zeta$ are the running couplings   which parametrize the model,  with  the indices $\alpha$ and $\beta$ running over   $1...D$.  This is, up  to a redefinition of the couplings,  the action used in \cite{paczuski88}  to investigate the crumpling transition.    Let us  recall  the physics  encoded  in Eq.(\ref{action}) at the mean-field level with   $u>0$ and $u+v D>0$.  For   $\zeta^2=0$,   the minimum of  $\Gamma_k$ is given by a configuration  where $\partial_{\alpha} \bf  r$  vanishes which  characterizes  a crumpled phase.  For    $\zeta^2>0$  this minimum  is given by a configuration $ {\bf r}({\bf x})=\zeta\sum_{\alpha=1} ^D x_{\alpha}\:\bf{e}_{\alpha}$, 
 where the $\{\bf{e}_{\alpha}\}$  are  $D$ orthonormal vectors,   which corresponds to  a $D$-dimensional flat phase.  Action  (\ref{action})   thus  describes  a   transition between a high-temperature, crumpling, phase and a low-temperature, flat, phase.    The excitation spectrum in the ordered phase is provided by  $d-D$ out-of-plane, capillary,  waves  and   $D$  in-plane, phonon, modes.  A  crucial  aspect of our  approach is that,  since  we establish  nonperturbative RG equations for the couplings  entering in Eq.(\ref{action}), and in particular   for the  coupling $\zeta$, we   are   able to tackle   both the crumpling transition,   typically associated to a  vanishing  $\zeta$,  and  the  flat phase fixed point (FLFP)   which is reached  by  letting   $\zeta$ run  to infinity.

  Technically,  the  flow equations for the couplings  ${Z, u,v}$  and $\zeta$  are obtained   using  their   definitions  in terms of functional  derivatives of the effective action (see \cite{berges02, delamotte03,kownacki09} for details)  and applying  RG  Eq.(\ref{renorm}).   In terms of dimensionless quantities, these equations  write as follows:
\begin{eqnarray}
\partial_{t}\zeta^2 & = & -(D-2+\eta_t)\,\zeta^2+\frac{4\, A_{D}}{D}\,\big\{  (D-1)\,\frac{(2u+vD)}{u+vD}\, l_{010}^{D+2}\nonumber\\
 & +&\frac{3u+(D+2)v}{u+vD} l_{001}^{D+2}+(d-D)l_{100}^{D+2}\big\}\nonumber\\
\partial_{t}u & = & (D-4+2\eta_t)u+\frac{16\, A_{D}}{D(D+2)}\left\{ 2\,(3u+2v)^{2}\, l_{002}^{D+4}\right.\nonumber\\ 
&+&\left.4D\, u(u+v)\, l_{011}^{D+4}+u^{2}(D^{2}+2D-8)\, l_{020}^{D+4}\right.\nonumber\\
& +& \left.2u^{2}(d-D)\, l_{200}^{D+4}\right\}\label{RG} \\
\partial_{t}v & = & (D-4+2\eta_t)v+\frac{16\, A_{D}}{D(D+2)}\big\{ -4u(u+v)\, l_{011}^{D+4}\nonumber\\
 & +&(d-D)\big(u^{2}+2(D+2)uv+D(D+2)v^{2}\big)l_{200}^{D+4}\nonumber\\
 &+&\big((3D+2)u^{2}+(D^{2}+D-2)(4uv +Dv^2)\big)l_{020}^{D+4}\nonumber\\
 &+&\big(9u^{2}+6(D+4)uv+(D^{2}+6D+12)v^{2}\big)l_{002}^{D+4}\big\} \nonumber
\end{eqnarray}
where $A_D=2^{-D-1}\pi^{-D/2}/\Gamma(D/2)$. The  flow of $Z$,  that provides the  function $\eta_t=- d\ln Z/dt$ giving  the critical exponent $\eta$ at a FP,  is too  long to be displayed here  (see \cite{kownacki09}).  In Eqs.(\ref{RG})   $l_{abc}^{D}$ is a shortcut   for :
\begin{equation}
\begin{array}{ll}
\displaystyle  l^D_{abc} = \frac{-1}{4\, A_{D}}\,\widehat{\frac{\partial}{\partial t}}\int d^{D}q\  \frac{1}{\left[P_0(q)\right]^a} \frac{1}{\left[P_1(q)\right]^b} \frac{1}{\left[P_2(q)\right]^c}
\label{threshold}
\end{array} 
\end{equation} 
where  $P_i(q)=Z\, q^{4}+R_{k}(q)+m_i^2\ q^{2}, i=0,1,2$  and $\widehat{{\partial}}/{\partial t}$ only acts on
$R_{k}$.  These  so-called "threshold  functions"  (see \cite{berges02,delamotte03})  control  the relative  role of the different   modes,    phonons and capillary waves,  within  the RG flow.  
In  Eq.(\ref{threshold}),    the mass   $m_0=0$ is associated to   the  $d-D$  transversal,  capillary, modes while  
$m_1^2\equiv 4 \zeta^2 u$ and $m_2^2\equiv 8\zeta^2 (u+v)$ are  masses associated to the $D$ phonons modes that split up  into  $D-1$ modes  with  mass $m_1$ and one mode with  mass $m_2$.

 {\sl -- The crumpling transition -- } Let us first  consider the crumpling transition.  To recover the  RG equations  derived  perturbatively  in \cite{paczuski88}   one  expands  Eq.(\ref{RG})    in powers of  both $\epsilon=4-D$ and   the couplings  $u$ and $v$ that are of order  $\epsilon$  at any putative  nontrivial    FP.  This also corresponds to an expansion in powers of the phonon masses, which are small at the crumpling transition FP. Using the fact that the  threshold functions     entering in the flow of $u$ and $v$ have a universal, cut-off independent,  limit  at  vanishing masses   in $D=4$ given by $l_{abc}^{8}=1$,  one obtains:
\begin{eqnarray}
\displaystyle\partial_ t u&=&\displaystyle-\epsilon u +{(d+21) u^2+20 v u+4 v^2\over 24 \pi ^2}
\label{RGeps}
\\
\displaystyle \partial_t v&=&\displaystyle -\epsilon  v+\frac{(d+15) u^2+4 (3 d+17) v u+4 (6 d+7) v^2}{48 \pi
   ^2}\nonumber . 
\end{eqnarray} 
Up to a change in  variable  ($v\to v-u/4$)  these are  the equations derived in \cite{paczuski88}.  
We recall that, at sufficiently high values  of $d$,  {\it i.e.}, $d> d_{cr}=219$,  just below $D=4$,  the   sets  of  Eqs.(\ref{RG})   and (\ref{RGeps})  admit  a   stable  (in the  $u$ and $v$ directions)  FP   associated to the crumpling transition,  called  crumpling transition fixed point (CTFP).   Still at $d>d_{cr}$,   there   exists another FP,  close to the CTFP,  which is unstable and that,   when the dimension $d$  is lowered   to $d_{cr}$,  annihilates with the CFTP,   defining  the curve $d_{cr}(D)$.  A  large-$d$ analysis of Eqs.(\ref{RG})  can be also easily done.   The leading contributions come from the capillary modes  which enter in  Eq.(\ref{RG})    through  the  terms  proportional to   $d-D$.  With our cut-off function  $l_{100}^{D}=4/D$ and $l_{200}^{D}=8/D$ so that the  coordinates of the   CTFP   are  given by  $\zeta^{2}_{cr}= 16  A_D/D( D^2-4)$, $u_{cr}=(16-D^2)D(2+D)/(256 d A_D)$ and $v_{cr}=-(16-D^2)D/(256 d A_D)$.   The corresponding critical exponents are:  $\nu=1/(D-2) +O(1/d)$  and $\eta=O(1/d)$ in agreement with \cite{david88}  and \cite{aronovitz89}.

\begin{figure}[htbp]
\vspace{-0.cm}
\includegraphics[width=0.9\linewidth,totalheight=55mm,clip,origin=tl]{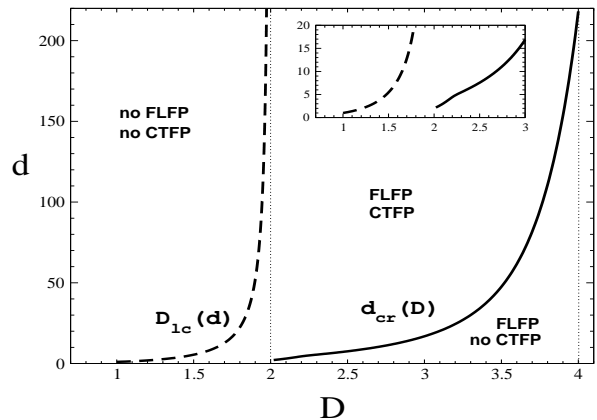}
\vspace{0cm}
\caption{On the right part,  the curve $d_{cr}(D)$ which  separates the region with a CTFP  and without a CTFP.  On the left part,  the lower critical dimension  $D_{lc}(d) $. }
\label{dD} 
\end{figure}

 To  tackle with the physics below $D=4$ we have numerically  solved the FP  equations   between $D=4$ and $D=2$, a dimension  in which the effects of truncation  start to be important. The right part of Fig.\ref{dD}  summarizes  our results: one finds  a smooth curve  $d_{cr}(D)$  which starts  at  $d_{cr}=219$ in $D=4$  and reaches   $d_{cr}\simeq 2$ in $D=2$  leading to predict  a second-order phase transition for physical membranes.  In this last case one finds, at  the CTFP,   a thermal exponent $\nu=0.52$ and $\eta=0.627$    which compares well  with the results  provided by the  large-$d$ expansion  $\eta=2/3$ \cite{david88,aronovitz89}  and  MC  results  $\eta=0.71(5)$   \cite{bowick96}  but   less  with the Monte Carlo Renormalization Group  $\eta=0.85 (15)$  \cite{espriu96} and  the  SCSA  $\eta=0.535$ \cite{ledoussal92}.  At our level of  approximation,
our results   display a weak dependence  with respect to the cut-off  function $R_k(q)$  that induces an error on the curve $d_{cr}(D)$. Using another  cut-off,  $R_k(q)=Z q^4/(\exp(q^4/k^4) -1)$,   we have evaluated    the error bar  on $d_{cr}(D=2)$, which is typically of order  $\delta d_{cr}\sim 1$.  This means that one cannot   exclude   $d_{cr}(D=2)$ to   be  close to,   or even  slightly  above,   $d=3$  so  that the crumpling transition for genuine  membranes   would  be predicted to be of  weak first-order in agreement with  recent MC results   \cite{kownacki02,koibuchi04}. This point will be  further analyzed in the near future  \cite{kownacki09}. 

{\sl -- The flat phase -- } The equations relevant to study the  flat phase are  easily  obtained  in our formalism  by considering the regime  $\zeta\gg 1$  in the RG  flow  Eqs.(\ref{RG}),  which   corresponds to a regime where the  phonon masses  are  very large and thus to a regime  dominated by   the  fluctuations of the  capillary waves, as expected in the deep flat phase.   Setting $\tilde d=d-D$ one gets: 
\begin{equation}
\hspace{-0.75cm}\partial_t u=(D-4+2 \eta_t) \ u+\displaystyle{256\  \tilde d\  u^2\ \tilde A_D\over D(D+2)(D+4)(D+8)}\nonumber
\end{equation} 
\vspace{-0.6cm}
\begin{equation}
\begin{array}{ll}
\hspace{-0.5cm}\partial_t v=(D-4+2 \eta_t) \ v\  +  \\
\vspace{-0.2cm}
\\ 
\vspace{-0.4cm}
\hspace{0.4cm}\displaystyle{128\  \tilde d\  (u^2+2(D+2)u v +D(D+2) v^2) \tilde  A_D \over D(D+2)(D+4)(D+8)}\label{flat}
 \end{array}
\end{equation} 
\vspace{-0.2cm}
\begin{equation}
\eta_t={128  (D+4)(D^2-1)u(u+2 v) A_D \over (D^4+6D^3+8D^2)(u+v)+128(D^2-1)u(u+2v)A_D}\nonumber
\end{equation} 
with $\tilde A_D=A_D(8+D-\eta_t)$ and, for  $\alpha=1/\zeta^2$:
\begin{equation}
\partial_t  \alpha=(D-2+ \eta_t) \alpha-\displaystyle{16\  \tilde d (6+D-\eta_t) \alpha^2 A_D\over D(D^2+8D+12) }   
\label{alpha}
\end{equation} 
an equation which generalizes, to any value of  $D$  and $d$,  the one  obtained  in the limit  of  large  elastic constants,  $D=2$  and large-$d$, in  \cite{david88}.   Note that the   function  $\eta_t$ in  Eqs.(\ref{flat})  and  (\ref{alpha}) determines, at a FP,    the exponent $\eta$  of the capillary waves.  The  analog  exponent, $\eta_u$, for the phonon modes,  is   obtained by the usual Ward identity \cite{aronovitz89}:  $\eta_u=4-D-2 \eta$ that follows from rotational invariance.

The  set of Eq.(\ref{flat}), when expanded in powers of $\epsilon=4-D$, degenerates  into those   derived   perturbatively  in  \cite{aronovitz88}.   Accordingly,  Eqs.(\ref{flat}-\ref{alpha}),    admit three nontrivial  FPs, among which one, the FLFP,   is stable with respect to  {\sl all}  directions including  $\alpha$ down to a dimension  $D_{lc}$,  the lower critical dimension.    In the limit  of  large codimension $\tilde d$   the  coordinates of the FLFP 
 are $\alpha_{f}=0$ ($\zeta^2_{f}\to\infty$),  $u_{f}=(16-D^2)D(2+D)/(256 d A_D)$ and $v_{f}=-(16-D^2)D/(256 d A_D)$,  with these two  last quantities being identical to those of the CTFP.   At the FLFP one finds   $\eta=O(1/d)$, in agreement with previous large-$d$  approach    \cite{aronovitz89}. Moreover  Eq.(\ref{alpha})  indicates that,  at  large $d$,   the FLFP      is stable down  to $D_{lc}(d\to\infty)=2$,  in agreement with \cite{aronovitz89}  which  predicts: $D_{lc}(d\to\infty)=2-2/d+O(1/d^2)$. Note also  that the RG flow on $u$ and $v$ indicates  that  for $\tilde d=0$ and at any nontrivial FP one has  the  exact result $\eta=(D-4)/2$   \cite{ledoussal92}.  For  physical membranes  one  finds $\eta=0.849$ which compares well with the SCSA $\eta=0.821$   \cite{ledoussal92} and numerical simulations   $\eta=0.750(5)$   \cite{bowick96} and  $\eta=0.81(3)$  \cite{zhang93}  but  less  with  the large-$d$ result   $\eta=2/3$  \cite{aronovitz89}.  
 
  Finally Eqs.(\ref{flat}-\ref{alpha})   also allow  a  determination of  $D_{lc}(d)$  for {\it all}  values of $d$. To do this we use  the equality \cite{aronovitz89}: $\eta(D_{lc},u_{f},v_{f})=2-D_{lc}$   at   the FLFP  which  defines  $D_{lc}$ as  the dimension   at  which  phonons and capillary waves scale  identically   \cite{aronovitz89}.     In fact,   the   RG  Eq.(\ref{alpha})   provides another interpretation of  $D_{lc}$. Indeed,  above $D_{lc}$,   Eq.(\ref{alpha})   possesses   a solution  with $\alpha\ne 0$  which  corresponds to   the  CTFP. Indeed, just above $D_{lc}$,   one has  $D-2+\eta_{cr}\ll 1 $   so that   the nontrivial solution of   Eq.(\ref{alpha})  obeys  $\alpha_{cr}\ll 1$ and thus   $\zeta_{cr}\gg 1$ which is precisely the regime of validity  of this equation. Thus, just  above $D_{lc}$,  Eq.(\ref{alpha})   well describes both the  CTFP and the FLFP which coexist and     $D_{lc}$  corresponds to  the dimension    at   which the CTFP  collapses  to the  FLFP which becomes unstable.   Using the relation  $\eta(D_{lc},u_{f},v_{f})=2-D_{lc}$ one  obtains:
 \begin{equation}
 d={D_{lc}^4+6 D_{lc}^3-3D_{lc}^2+4D_{lc}\over 2(-D_{lc}^2-D_{lc}+6)}\ 
 \end{equation}
  which, once  inversed,   provides  the expression of $D_{lc}(d)$. The corresponding curve is displayed on the left part of Fig.\ref{dD}.  In particular one has, for genuine membranes: $D_{lc}(d=3)\simeq 1.33$.  This result  displays a remarkable   stability  with respect to a change in  the cut-off function. With $R_k(q)=Z q^4/(\exp(q^4/k^4) -1)$  one finds $D_{lc}(d=3)\simeq 1.30$. Our results  compare well  with the  large-$d$, $D_{lc}(d=3)=4/3$ \cite{aronovitz89},  and SCSA,  $D_{lc}(d=3)= 1.5$ \cite{ledoussal92}.

 In summary,  we have investigated  the  crumpling transition and flat phase  of  $D$-dimensional polymerized membranes embedded in a $d$-dimensional  space  within  a nonperturbative  RG approach.   We have  determined the whole line $d_{cr}(D)$ that separates the second- and first-order  regions,  the lower critical dimension  $D_{lc}(d)$ and the critical exponents.   More sophisticated  ansatz should be used  to  systematically increase the accuracy of  our  results although  implying a heavy algebra.   Finally, our  approach can  be applied to   many  other situations in which  the perturbative approaches lead to an unsatisfying  quantitative or even qualitative description such as  in  self-avoiding, anisotropic or disordered membranes \cite{proceedings89,radzihovsky04,duplantier04}.

\acknowledgments{We thank F. David, B. Delamotte, T.C. Lubensky,  L. Radzihovsky and J. Vidal   for helpful discussions.}


\begin{thebibliography}{26}
\expandafter\ifx\csname natexlab\endcsname\relax\def\natexlab#1{#1}\fi
\expandafter\ifx\csname bibnamefont\endcsname\relax
  \def\bibnamefont#1{#1}\fi
\expandafter\ifx\csname bibfnamefont\endcsname\relax
  \def\bibfnamefont#1{#1}\fi
\expandafter\ifx\csname citenamefont\endcsname\relax
  \def\citenamefont#1{#1}\fi
\expandafter\ifx\csname url\endcsname\relax
  \def\url#1{\texttt{#1}}\fi
\expandafter\ifx\csname urlprefix\endcsname\relax\def\urlprefix{URL }\fi
\providecommand{\bibinfo}[2]{#2}
\providecommand{\eprint}[2][]{\url{#2}}

\bibitem[{\citenamefont{Nelson et~al.}(2004)\citenamefont{Nelson, Piran, and
  Weinberg}}]{proceedings89}
\bibinfo{editor}{\bibfnamefont{D.~R.} \bibnamefont{Nelson}},
  \bibinfo{editor}{\bibfnamefont{T.}~\bibnamefont{Piran}}, \bibnamefont{and}
  \bibinfo{editor}{\bibfnamefont{S.}~\bibnamefont{Weinberg}}, eds.,
  \emph{\bibinfo{title}{Proceedings of the Fifth Jerusalem Winter School for
  Theoretical Physics}} (\bibinfo{publisher}{World Scientific, Singapore},
  \bibinfo{year}{2004}), \bibinfo{edition}{2nd} ed.

\bibitem[{\citenamefont{{{See the contribution of L. Radzihovsky in
  [1]}}}()}]{radzihovsky04}
\bibinfo{author}{\bibnamefont{{{See the contribution of L. Radzihovsky in
  [1]}}}}.

\bibitem[{\citenamefont{Bowick and Travesset}(2001)}]{bowick01}
\bibinfo{author}{\bibfnamefont{M.~J.} \bibnamefont{Bowick}} \bibnamefont{and}
  \bibinfo{author}{\bibfnamefont{A.}~\bibnamefont{Travesset}},
  \bibinfo{journal}{Phys. Rep.} \textbf{\bibinfo{volume}{344}},
  \bibinfo{pages}{255} (\bibinfo{year}{2001}).

\bibitem[{\citenamefont{{{See the contributions of D. R. Nelson in
  [1]}}}()}]{nelson04}
\bibinfo{author}{\bibnamefont{{{See the contributions of D. R. Nelson in
  [1]}}}}.

\bibitem[{\citenamefont{Nelson and Peliti}(1987)}]{nelson87}
\bibinfo{author}{\bibfnamefont{D.~R.} \bibnamefont{Nelson}} \bibnamefont{and}
  \bibinfo{author}{\bibfnamefont{L.}~\bibnamefont{Peliti}},
  \bibinfo{journal}{J. Phys. (Paris)} \textbf{\bibinfo{volume}{48}},
  \bibinfo{pages}{1085} (\bibinfo{year}{1987}).

\bibitem[{\citenamefont{David and Guitter}(1988)}]{david88}
\bibinfo{author}{\bibfnamefont{F.}~\bibnamefont{David}} \bibnamefont{and}
  \bibinfo{author}{\bibfnamefont{E.}~\bibnamefont{Guitter}},
  \bibinfo{journal}{Europhys. Lett.} \textbf{\bibinfo{volume}{5}},
  \bibinfo{pages}{709} (\bibinfo{year}{1988}).

\bibitem[{\citenamefont{Paczuski et~al.}(1988)\citenamefont{Paczuski, Kardar,
  and Nelson}}]{paczuski88}
\bibinfo{author}{\bibfnamefont{M.}~\bibnamefont{Paczuski}},
  \bibinfo{author}{\bibfnamefont{M.}~\bibnamefont{Kardar}}, \bibnamefont{and}
  \bibinfo{author}{\bibfnamefont{D.~R.} \bibnamefont{Nelson}},
  \bibinfo{journal}{Phys. Rev. Lett.} \textbf{\bibinfo{volume}{60}},
  \bibinfo{pages}{2638} (\bibinfo{year}{1988}).

\bibitem[{\citenamefont{Aronovitz and Lubensky}(1988)}]{aronovitz88}
\bibinfo{author}{\bibfnamefont{J.~A.} \bibnamefont{Aronovitz}}
  \bibnamefont{and} \bibinfo{author}{\bibfnamefont{T.~C.}
  \bibnamefont{Lubensky}}, \bibinfo{journal}{Phys. Rev. Lett.}
  \textbf{\bibinfo{volume}{60}}, \bibinfo{pages}{2634} (\bibinfo{year}{1988}).

\bibitem[{\citenamefont{Guitter et~al.}(1989)\citenamefont{Guitter, David,
  Leibler, and Peliti}}]{guitter89}
\bibinfo{author}{\bibfnamefont{E.}~\bibnamefont{Guitter}},
  \bibinfo{author}{\bibfnamefont{F.}~\bibnamefont{David}},
  \bibinfo{author}{\bibfnamefont{S.}~\bibnamefont{Leibler}}, \bibnamefont{and}
  \bibinfo{author}{\bibfnamefont{L.}~\bibnamefont{Peliti}},
  \bibinfo{journal}{J. Phys. (Paris)} \textbf{\bibinfo{volume}{50}},
  \bibinfo{pages}{1787} (\bibinfo{year}{1989}).

\bibitem[{\citenamefont{Aronovitz et~al.}(1989)\citenamefont{Aronovitz,
  Golubovic, and Lubensky}}]{aronovitz89}
\bibinfo{author}{\bibfnamefont{J.~A.} \bibnamefont{Aronovitz}},
  \bibinfo{author}{\bibfnamefont{L.}~\bibnamefont{Golubovic}},
  \bibnamefont{and} \bibinfo{author}{\bibfnamefont{T.~C.}
  \bibnamefont{Lubensky}}, \bibinfo{journal}{J. Phys. (Paris)}
  \textbf{\bibinfo{volume}{50}}, \bibinfo{pages}{609} (\bibinfo{year}{1989}).

\bibitem[{\citenamefont{{{See the contribution of Y. Kantor in
  [1]}}}()}]{kantor04}
\bibinfo{author}{\bibnamefont{{{See the contribution of Y. Kantor in [1]}}}}.

\bibitem[{\citenamefont{{{See the contribution of G. Gompper and D. M. Kroll in
  [1]}}}()}]{gompper04}
\bibinfo{author}{\bibnamefont{{{See the contribution of G. Gompper and D. M.
  Kroll in [1]}}}}.

\bibitem[{\citenamefont{{{J. -Ph. } Kownacki} and Diep}(2002)}]{kownacki02}
\bibinfo{author}{\bibnamefont{{{J. -Ph. } Kownacki}}} \bibnamefont{and}
  \bibinfo{author}{\bibfnamefont{H.~T.} \bibnamefont{Diep}},
  \bibinfo{journal}{Phys. Rev. E} \textbf{\bibinfo{volume}{66}},
  \bibinfo{pages}{066105} (\bibinfo{year}{2002}).

\bibitem[{\citenamefont{{{H. Koibuchi et. al.}}}(2004)}]{koibuchi04}
\bibinfo{author}{\bibnamefont{{{H. Koibuchi et. al.}}}},
  \bibinfo{journal}{Phys. Rev. E} \textbf{\bibinfo{volume}{69}},
  \bibinfo{pages}{066139} (\bibinfo{year}{2004}).

\bibitem[{\citenamefont{Paczuski and Kardar}(1989)}]{paczuski89}
\bibinfo{author}{\bibfnamefont{M.}~\bibnamefont{Paczuski}} \bibnamefont{and}
  \bibinfo{author}{\bibfnamefont{M.}~\bibnamefont{Kardar}},
  \bibinfo{journal}{Phys. Rev. A} \textbf{\bibinfo{volume}{39}},
  \bibinfo{pages}{6086} (\bibinfo{year}{1989}).

\bibitem[{\citenamefont{{Le Doussal} and Radzihovsky}(1992)}]{ledoussal92}
\bibinfo{author}{\bibfnamefont{P.}~\bibnamefont{{Le Doussal}}}
  \bibnamefont{and}
  \bibinfo{author}{\bibfnamefont{L.}~\bibnamefont{Radzihovsky}},
  \bibinfo{journal}{Phys. Rev. lett.} \textbf{\bibinfo{volume}{69}},
  \bibinfo{pages}{1209} (\bibinfo{year}{1992}).

\bibitem[{\citenamefont{Wetterich}(1993)}]{wetterich93c}
\bibinfo{author}{\bibfnamefont{C.}~\bibnamefont{Wetterich}},
  \bibinfo{journal}{Phys. Lett. B} \textbf{\bibinfo{volume}{301}},
  \bibinfo{pages}{90} (\bibinfo{year}{1993}).

\bibitem[{\citenamefont{Bagnuls and Bervillier}(2001)}]{bagnuls01}
\bibinfo{author}{\bibfnamefont{C.}~\bibnamefont{Bagnuls}} \bibnamefont{and}
  \bibinfo{author}{\bibfnamefont{C.}~\bibnamefont{Bervillier}},
  \bibinfo{journal}{Phys. Rep.} \textbf{\bibinfo{volume}{348}},
  \bibinfo{pages}{91} (\bibinfo{year}{2001}).

\bibitem[{\citenamefont{Berges et~al.}(2002)\citenamefont{Berges, Tetradis, and
  Wetterich}}]{berges02}
\bibinfo{author}{\bibfnamefont{J.}~\bibnamefont{Berges}},
  \bibinfo{author}{\bibfnamefont{N.}~\bibnamefont{Tetradis}}, \bibnamefont{and}
  \bibinfo{author}{\bibfnamefont{C.}~\bibnamefont{Wetterich}},
  \bibinfo{journal}{Phys. Rep.} \textbf{\bibinfo{volume}{363}},
  \bibinfo{pages}{223} (\bibinfo{year}{2002}).

\bibitem[{\citenamefont{Delamotte et~al.}(2004)\citenamefont{Delamotte,
  Mouhanna, and Tissier}}]{delamotte03}
\bibinfo{author}{\bibfnamefont{B.}~\bibnamefont{Delamotte}},
  \bibinfo{author}{\bibfnamefont{D.}~\bibnamefont{Mouhanna}}, \bibnamefont{and}
  \bibinfo{author}{\bibfnamefont{M.}~\bibnamefont{Tissier}},
  \bibinfo{journal}{Phys. Rev. B} \textbf{\bibinfo{volume}{69}},
  \bibinfo{pages}{134413} (\bibinfo{year}{2004}).

\bibitem[{\citenamefont{Litim}(2002)}]{litim02}
\bibinfo{author}{\bibfnamefont{D.~F.} \bibnamefont{Litim}},
  \bibinfo{journal}{Nucl. Phys. B} \textbf{\bibinfo{volume}{631}},
  \bibinfo{pages}{128} (\bibinfo{year}{2002}).

\bibitem[{\citenamefont{{{J. -P. } Kownacki} and Mouhanna}()}]{kownacki09}
\bibinfo{author}{\bibnamefont{{{J. -P. } Kownacki}}} \bibnamefont{and}
  \bibinfo{author}{\bibfnamefont{D.}~\bibnamefont{Mouhanna}},
  \bibinfo{note}{unpublished}.

\bibitem[{\citenamefont{{{M. J. Bowick et. al.}}}(1996)}]{bowick96}
\bibinfo{author}{\bibnamefont{{{M. J. Bowick et. al.}}}}, \bibinfo{journal}{J.
  Phys. (France) I} \textbf{\bibinfo{volume}{6}}, \bibinfo{pages}{1321}
  (\bibinfo{year}{1996}).

\bibitem[{\citenamefont{Espriu and Travesset}(1996)}]{espriu96}
\bibinfo{author}{\bibfnamefont{D.}~\bibnamefont{Espriu}} \bibnamefont{and}
  \bibinfo{author}{\bibfnamefont{A.}~\bibnamefont{Travesset}},
  \bibinfo{journal}{Nucl. Phys. B} \textbf{\bibinfo{volume}{468}},
  \bibinfo{pages}{514} (\bibinfo{year}{1996}).

\bibitem[{\citenamefont{Zhang et~al.}(1993)\citenamefont{Zhang, Davis, and
  Kroll}}]{zhang93}
\bibinfo{author}{\bibfnamefont{Z.}~\bibnamefont{Zhang}},
  \bibinfo{author}{\bibfnamefont{H.~T.} \bibnamefont{Davis}}, \bibnamefont{and}
  \bibinfo{author}{\bibfnamefont{D.~M.} \bibnamefont{Kroll}},
  \bibinfo{journal}{Phys. Rev. E} \textbf{\bibinfo{volume}{48}},
  \bibinfo{pages}{{R651}} (\bibinfo{year}{1993}).

\bibitem[{\citenamefont{{{See the contributions of B. Duplantier in
  [1]}}}()}]{duplantier04}
\bibinfo{author}{\bibnamefont{{{See the contributions of B. Duplantier in
  [1]}}}}.

\end{thebibliography}
\end{document}